\documentclass[10pt,english,preprint2]{aastex}
\usepackage[T1]{fontenc}
\usepackage[latin1]{inputenc}
\usepackage{graphicx}
\usepackage{amssymb}

\makeatletter

\usepackage{graphicx}
\usepackage{times}
\usepackage{courier}
\usepackage{babel}
\makeatother
\begin{document}

\title{Solar Magnetic Tracking. I. Software Comparison and Recommended Practices \small{\emph{(In press, \apj, 2007)}}}

\author{C.E. DeForest}

\affil{Southwest Research Institute, 1050 Walnut Street Suite 400, Boulder,
CO 80302}

\email{deforest@boulder.swri.edu, }

\author{H.J. Hagenaar}

\affil{Lockheed Martin Advanced Technology Center, Org. ADBS., Bldg. 252,
Palo Alto, CA 94304}

\email{hagenaar@lmsal.com}

\author{D.A. Lamb}

\affil{Dept. of Astrophysical and Planetary Science, University of Colorado,
Boulder 80309-0391 USA}

\email{derek@boulder.swri.edu}

\author{C.E. Parnell}

\affil{School of Mathematics and Statistics, St. Andrews University, St.
Andrews, Scotland KY16 9S}

\email{clare@mcs.st-and.ac.uk}

\and{ }

\author{B.T. Welsch}

\affil{University of California at Berkeley Space Sciences Laboratory, 7
Gauss Way, UCB, CA 94720-7450, USA}

\email{welsch@ssl.Berkeley.edu}

\begin{abstract}
Feature tracking and recognition are increasingly common tools for
data analysis, but are typically implemented on an ad-hoc basis by
individual research groups, limiting the usefulness of derived results
when selection effects and algorithmic differences are not controlled.
Specific results that are affected include the solar magnetic turnover
time, the distributions of sizes, strengths, and lifetimes of magnetic
features, and the physics of both small scale flux emergence and the
small-scale dynamo. In this paper, we present the results of a detailed
comparison between four tracking codes applied to a single set of
data from SOHO/MDI, describe the interplay between desired tracking
behavior and parameterization tracking algorithms, and make recommendations
for feature selection and tracking practice in future work.

\newpage
~\newpage

\end{abstract}


\newpage
\section{Introduction}

The last decade has seen a sea change in the way that solar physics
is accomplished. Advances in detector technology have permitted missions
such as SOHO (e.g. \citealt{Scherrer1995}) and TRACE (\citealt{Handy1999}),
and ground-based observatories such as GONG (\citet{Leibacher1995}),
to produce far more data than can be analyzed directly by humans.
The planned SDO mission (\citealt{Schwer2002}) will produce data
a thousand times faster. Hence, automated data mining has become a
necessary tool of the analysis trade. Applied to image data, data
mining consists of algorithmic recognition of visual features in the
data. Applications such as feature and pattern recognition fall within
the field of \emph{computer vision}, which is the subject of active
research in the computer science community.

Magnetic feature identification and tracking have proven useful for
extracting statistical parameters of the solar dynamo
(e.g. \citealt{Hagenaar1999,Schrijver1997}), allowing more
sophisticated analyses than have been possible by hand
(e.g. \citealt{HarveyKL1993}). Current applications of feature
tracking include characterization of bulk field behavior at the
photosphere, probing of the solar dynamo, identification of the
magnetic roots of solar atmospheric features, and constraint of MHD
models. Each of these applications is discussed below.

In the last few years, each of our research groups has independently
developed four separate tracking codes adapted to studying slightly
different aspects of the solar magnetic field.  CURV was the first
code developed to study magnetic features in the MDI quiet sun data
(\citealt{Hagenaar1999}); MCAT has been used to study interaction
between network flux elements (\citealt{Parnell2002}); SWAMIS
(\citealt{LambDeForest2003}) is intended to drive semi-empirical MHD
models of the quiet sun; and YAFTA (\citealt{Welsch2003}) was
developed to study active region dynamics.

Our separate tracking codes are similar enough to be applied to
similar problems and to yield directly comparable results.  However,
feature tracking is not a simple endeavor, and many subtle
characteristics of each code can strongly affect derived results.
This, together with the ad-hoc manner in which each tracking code
was developed, made it difficult to compare or duplicate results
between groups.

In 2004 November we met at St. Andrews University to reconcile results
from all four sets of software. We applied each code to a sample data
set and compared results from the different algorithms to reconcile
the results across research groups. Furthermore, we identified how
algorithmic choices affect magnetic feature tracking results, and
developed a set of recommended practices to guide future development
of feature tracking and related software for the solar community.

Developing a baseline of best recommended practices for feature
tracking and computer vision is an important goal for the solar
imaging community, because feature tracking is a fundamental component
of many types of data analysis. Applied to the solar magnetic field,
it has been used to characterize the statistical parameters of the
field by determining the distribution of feature sizes
and fluxes
(\citealt{HarveyKL1993,Hagenaar1999,Hagenaar2001,Parnell2002}) and the
average lifetime of individual features (\citealt{Hagenaar2003}).
Automated extraction of parameters such as clustering distributions
(\citealt{LambDeForest2003}) and event distributions
(\citealt{DeForestLamb2004}) are being used to derive more detailed
information about the solar dynamo. All of these applications are
dominated by the relationship between small scale event detections and
the noise floor of the instrument used for detection, generally a
line-of-sight / scalar magnetograph such as SOHO/MDI
(\citealt{Scherrer1995}) or GONG (\citealt{Leibacher1999}).

Feature tracking is further useful for constraining the energy input into flux
systems in the solar corona. Much of the energy deposited into the
chromosphere and corona is thought to be transported by the Poynting
vector, as photospheric motions do work on the magnetic field by
pushing magnetic flux around the surface
(e.g. \citealt{Parker1988,Fossum2004}).  Feature tracking allows
simple derivation of the motion field from time series of
images. \citet*{Welsch2004} used feature tracking to estimate the
quiet sun helicity flux into the corona, and \citet{DeForestLamb2004}
and Parnell (\citeyear{ParnellJupp2000,Parnell2002}) are using feature
tracking to identify the roots and nature of small scale heating
events such as bright points.

A third important application of feature tracking is to drive boundary
conditions of semi-empirical MHD models of the solar atmosphere, such
as are anticipated for space weather prediction. Time-dependent MHD
modeling requires knowledge not just of the three-dimensional vector
field at the surface of the Sun, but also of the motion of individual
lines of magnetic flux; feature tracking derives the motion
information from time series measurements of the magnetic
field. Indeed, Peano's existence and completeness theorem (see, e.g.,
\citealt{Simmons}) implies that knowledge of the initial magnetic
topology in the force-free upper layers of the atmosphere, together
with the radial component of the field at the lower boundary, is
equivalent to knowledge of the full vector field everywhere on the
lower boundary. Provided that the initial topology may be estimated,
this equivalence makes feature tracking a powerful tool for modeling
energy input into the solar atmosphere even in the absence of full
vector field measurements, as the distribution of radial magnetic flux
on the $\tau=1$ surface at the photosphere approximates the
distribution at the $\beta=1$ surface in the upper chromosphere.

In this article, the first in a series on results from tracking of
photospheric magnetic features, we discuss the state of the art and some
current applications of magnetic tracking software. In
\S\ref{sec:Discussion-of-Tracking} we outline the basic steps of a
feature tracking algorithm; in
\S\S\ref{sec:Tracking-Results:-A}-\ref{sec:Discussion} we present and
discuss the differences between the codes' results as applied to a
reference data set; and in \S\ref{sec:recommendations} we recommend
{}``best practices'' for future codes to follow for feature
tracking applications. Finally, \S\ref{sec:Conclusions} contains some
general conclusions and insights, and a glossary at the end contains
recommended vocabulary to describe specific aspects of magnetic
tracking.

\section{\label{sec:Discussion-of-Tracking}Discussion of Tracking Algorithms}

Feature tracking can be divided into five separate operations: (i)
image preprocessing; (ii) discrimination/detection; (iii) feature
identification within a frame; (iv) feature association across frames;
and (v) event detection. In addition, some noise filtering is
accomplished by filtering the associated features to discard
short-lived or small features that have too high a likelihood of being
noise. Here, we discuss the important components of magnetic feature
tracking algorithms in general, and outline the differences between
each of the four principal codes that we compared.

\subsection{Preprocessing}

In general, magnetograms arrive from an instrument with some level of
background noise and with position-dependent foreshortening due to the
curvature of the Sun.  Reducing the noise floor and eliminating
perspective effects requires preprocessing images before applying
feature recognition. Temporal averaging, projection angle scaling, and
resampling to remove perspective and solar rotation effects are
commonly applied before most high-level analysis. In particular, 0.5-2
arcsecond scale magnetograms benefit from being averaged over 5-12
minutes to reduce background noise, and line-of-sight (Stokes V)
magnetograms of the quiet sun benefit from being divided by a cosine
factor to account for the difference between the magnetogram line of
sight and local vertical at the surface of the Sun, under the model
that weak field is close to vertical at the photosphere. 

Spaceborne magnetographs such as MDI and the anticipated SDO are also
susceptible to cosmic ray spikes, which must be removed either by 
temporal filtering of tracked features or by preprocessing the images.

Most magnetograms made with a filtergraph type instrument such as MDI
or GONG contain at least three sources of random noise at each pixel:
(i) photon statistics, which produce a familiar white noise spectrum;
(ii) P-mode contamination, which is due to the five-minute Doppler
oscillations leaking into the Zeeman signal; and (iii) granulation
noise, which is due to solar evolution between the different
filtergraph exposures that make up each magnetogram. The photon shot
noise is a uniform random variable with an independent sample at every
pixel and a presumed Gaussian distribution. The P-mode contamination
is a random variable with far fewer independent spatial samples per
image, because of the low spatial frequencies of the P-modes, and an
oscillating temporal component. Granulation-based noise has a spatial
scale of a few arc seconds and a coherence time of 5 minutes. MDI is
well tuned so that the three sources of noise are about equal in
individual images; but in spatially binned, temporally averaged, or
smoothed images, the granulation and P-modes dominate the noise
spectrum.

Although data preprocessing is not part of the process of feature
identification and tracking, preprocessing effects can affect tracking
results and we recommend (in \S\ref{sec:recommendations}) specific 
practices to reduce artifacts.

\subsection{Discrimination}

Any feature-recognition algorithm requires \emph{discrimination},
i.e. the separation of foreground features from background noise.
Every magnetogram sequence appears to contain many faint features
at or slightly below the level of the noise floor, so discrimination
is not trivial.

The simplest discrimination scheme, direct thresholding, works well
only for strong magnetic features that are well separated from the
noise floor, such as flux concentrations in the magnetic network or
in active regions. Other types of magnetic feature, such as weak intranetwork
fields, suffer because keeping the threshold high enough to avoid
false-positive detections creates a large number of false-negative
non-detections of the weak magnetic features. The problem is the huge
number of individual detection operations (one per pixel per time
step), which makes false positives a significant problem. With a Gaussian
noise distribution, setting the threshold to three standard deviation
($\sigma$) units yields a false positive rate of about $10^{-4}$,
so that a 300-frame data set with dimension 300x300 pixels would yield
around three thousand false positive detections from noise alone,
and perhaps 10-30 times that number of inconsistently detected weak
features (false negatives). 

Each of our codes used a different discrimination scheme, affecting
what types of feature could be detected.  YAFTA, originally intended
for use with active region magnetograms well above the noise floor,
uses a simple threshold test to discriminate.  The other three codes
have adopted two different schemes to work closer to the noise floor,
both of which add additional tests to the basic threshold test.

SWAMIS and MCAT use hysteresis, used by \citet{LambDeForest2003} and
by \citet{Parnell2002}, in which two thresholds are applied: a high
threshold for isolated pixels, and a second, lower threshold for
pixels that are adjacent to already-selected pixels.  Adjacency is
allowed in space and/or time. The hysteresis misses some very weak
features, but captures every feature that at some location and/or time
exceeds the large threshold. The proximity requirement reduces the
number of pixels that undergo the lower threshold test, and therefore
reduces the number of false-positive detections.  Depending on
application, the higher threshold is chosen to be $3-6\,\sigma$, and
the lower threshold $1-3\,\sigma$, where $\sigma^{2}$ is the variance
(and $\sigma$ is the RMS variation) of the data.

MCAT and SWAMIS differ slightly in the nature of the hysteresis.  Both
codes use separate masks for positive and negative flux
concentrations, but MCAT applies the low threshold to any pixels in
the current frame that are adjacent to a detected feature in the next
or previous frame.  MCAT makes one forward pass through the data,
comparing pixels in each frame to the higher threshold at most
locations and to the lower threshold in locations that were occupied
in the previous frame; and then one reverse pass that is identical
except that the low-threshold mask comes from the next, rather than
previous, frame.  

SWAMIS uses a ``contagion'' algorithm that treats the time axis as a
third spatial dimension: a pixel is subjected to the low threshold if
it is adjacent, either in time or in space, to any detected pixel.
Considering pixels as cubes in (x,y,t) space, each pixel is subjected
to the lower threshold if it shares at least one edge with a pixel
that has been marked occupied and that has the same sign.  The
contagion algorithm is executed in a single pass through the data with
in-frame recursion to dilate the detected regions annd with
backtracking to re-test newly ``infected'' pixels in previous frames.

CURV uses the curvature method used by \textbf{}\citet{Strous1996} and
by \citet{Hagenaar1999}, in which both the data
values and their second derivative are tested. To be considered part
of a local maximum/minimum by the curvature algorithm, the magnitude
of a pixel must exceed a value threshold and all surrounding pixels
must have a negative/positive second derivative in each of the
horizontal, vertical, and two diagonal directions. The second
derivative criterion adds four additional independent threshold tests
for each pixel, reducing the number of false detections at a given
threshold and allowing single threshold values comparable to the low
threshold values used in SWAMIS and MCAT.

All three of MCAT, SWAMIS, and CURV impose minimum-size and lifetime
requirements on features at a later step in the processing, reducing the
effect of false positives in the detection step.  YAFTA also imposes a 
minimum lifetime requirement to reduce false positives from noise 
fluctuation.

\subsection{\label{sub:Feature-identification}Feature identification}

Feature identification is the operation of connecting masked pixels
into distinct identifiable (and identified) structures in each frame.
In practice, this means forming a detected feature map, an image whose
pixels have integer numeric values that correspond to index numbers of
particular features. Each of our codes uses a variant of a clumping
dilation algorithm that identifies connected loci of pixels within a
masked region.

\begin{figure}[!tb]
\begin{center}\center{\includegraphics[%
  width=3in,
  height=1.14in]{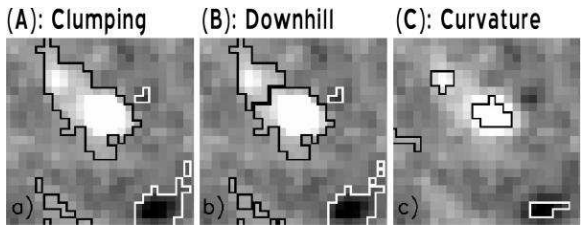}}\end{center}

\caption{\label{fig:identification-types}Effect of different feature-identification
schemes on the identified structure of a large flux concentration.
(A) Clumping identifies all connected above-threshold pixels into
a single feature. (B) Downhill methods identify one feature per local
maximum region. (C) Curvature methods identify the convex core around
each local maximum.}
\end{figure}

MCAT clumps masked pixels directly into contiguous regions.
YAFTA and SWAMIS can switch between direct clumping and a gradient
based (``downhill'') method that dilates local maxima by expansion
down the gradient toward zero flux density. CURV also uses direct
clumping, but generates initial feature masks with a data-value curvature
method that restricts the features to isolated regions, yielding features
that are segmented more like those of the downhill method in YAFTA
and SWAMIS than like the other clumping codes. All three techniques
are illustrated in Figure \ref{fig:identification-types}.

The tradeoff between the clumping and downhill methods is that the
downhill method is better at picking out the structure of individual
clusters of magnetic flux, while the clumping method is somewhat less
noise-susceptible. Fluctuations from either solar convection or
instrument noise can easily create small local maxima that are
identified as transient structures by the downhill method; simple
clumping eliminates these small transients, for better or for worse.

Which method is appropriate depends on the specific scientific
application, as discussed further in \S
\ref{sec:recommendations}.

\subsection{\label{sub:Feature-association}Feature association}

\begin{figure}[!tb]
\center{\includegraphics[%
  width=3in,
  height=2.473in]{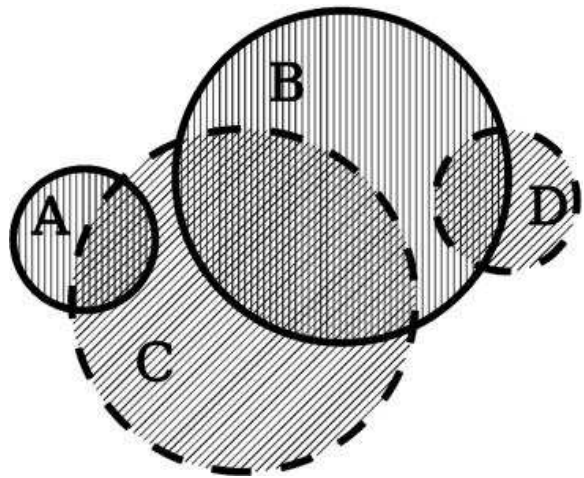}}

\caption{\label{fig:assoc}A pathological association case. Features A and
B are in the previous frame, C and D in the current frame. A maximum-overlap
method associates B and C. The recommended associative algorithm (dual-maximum
overlap) associates B=C if and only if $B\cap C$ is the largest of
C's intersecting regions and also the largest of B's intersecting
regions. A and B merge to form C, at the same time that D calves via
fragmentation from B.}
\end{figure}

Feature association is the fixing of a feature's identity across
different frames of an image sequence. Most features in adjacent
frames of an image are related by similarity of position and shape:
when a feature in frame $m+1$ is sufficiently similar to the feature
in frame $m$, then it is likely that the two features represent the
same physical object at the two different times. All of our existing
codes use variations of a dual-maximum-overlap criterion to identify
persistent features across frames (Figure \ref{fig:assoc}); this
technique associates two features B and C in adjacent frames only if
$B\cap C$ is larger (in a flux-weighted sense) than any other
intersection with either $B$ or $C$.

All four of our codes follow variants of the largest-intersection
criterion, either following maximum flux overlap or maximum area
overlap.  YAFTA uses arbitrary label choice on its first pass through
the data, and then ``label conflicts'' in a subsequent pass that
uses maximum overlap.

\subsection{Filtering based on size/longevity}

When working close to the noise floor, it is useful to reject small
features, because false positives are much more likely in small
clusters of pixels than in large ones. All of our codes reject
identified features that do not meet some minimum size
criterion. Criteria that are useful include: maximum size; average
size; lifetime; or total number of pixels across the life of the
feature. The filtering can be accomplished only after feature
identification (for per-frame size checks) or feature association (for
maximum size checks or longevity checks). 

Additional problems exist due to fluctuations in background noise that
may cause features that appear for only a single frame, or may cause weak
but persistent features to disappear for a frame (the \emph{Swiss
cheese problem}).  Similarly, associated features may split and then
re-merge rapidly due to fluctuations in a single frame (the
\emph{oscillating twins} problem).

CURV sidesteps association problems by requiring oversampling on the
time axis of the input data cube; this reduces the frame-to-frame
fluctuations of individual features.

MCAT avoids both problems with a three-step process.  (i)
completely-surrounded holes in the center of a feature are filled in
and the missing pixels are counted as part of the feature; (ii) Twin
features that merge for a single frame are forced to remain separate;
(iii) single features that split for a single frame are forced to
remain merged.

SWAMIS overcomes both the oscillating twins and Swiss cheese problems
by re-associating short lived features with nearby larger features if
there is sufficient overlap between them.

\subsection{Classification of Origin and Demise}

Identifying and locating individual features as they evolve is
properly described as \emph{feature tracking}, but identifying
structures and events that may include several features is more
properly described as a complete \emph{computer vision}
application. Not all of our feature tracking codes include provision
to detect and identify interactions of multiple flux concentrations,
such as pairwise emergence, but such detection is an important part of
characterizing magnetic evolution and hence is discussed here. In
particular, because magnetic features are not corks but rather
cross-sections of curvilinear manifolds (field lines that pass through
the photosphere), they are connected pairwise by the magnetic
field. Identifying the association between freshly emerged pairs thus
gives useful information about the overall field topology and how it
changes via reconnection of the overlying field before the death
(e.g. by submergence) of the individual features.

The origin and demise of features is different than the origin and
demise of magnetic flux itself: in particular, features can fragment
or merge under the influence of the photospheric flow field (e.g.
\citealt{Schrijver1997}) without any flux emerging or submerging through
the photospheric surface. Fragmentation and merging can result in
apparent violation of the conservation of flux as magnetic flux sinks
below or rises above the detection threshold of the instrument being
used to detect it.

Software to identify origin events recognizes flux concentrations
near each newly detected concentration, and classifies the origin
according to these nearby concentrations and the time derivative of
their contained flux. To avoid missing associated structure, an allowed
margin of error is required in the spatial or temporal offset between
two associated features, and also in the flux rate-of-change between
the features. 

Demise events are similar to origin events and may be recognized with
the same code, operating on tracked data in reverse time order. As
with birth events, demise events are not necessarily related to emergence
or submergence of magnetic flux.

\section{\label{sec:Tracking-Results:-A}Tracking Results: A Comparison Across
Codes}

\begin{figure}[!tbh]
\begin{center}\center{\includegraphics[%
  width=3in,height=5.9in]{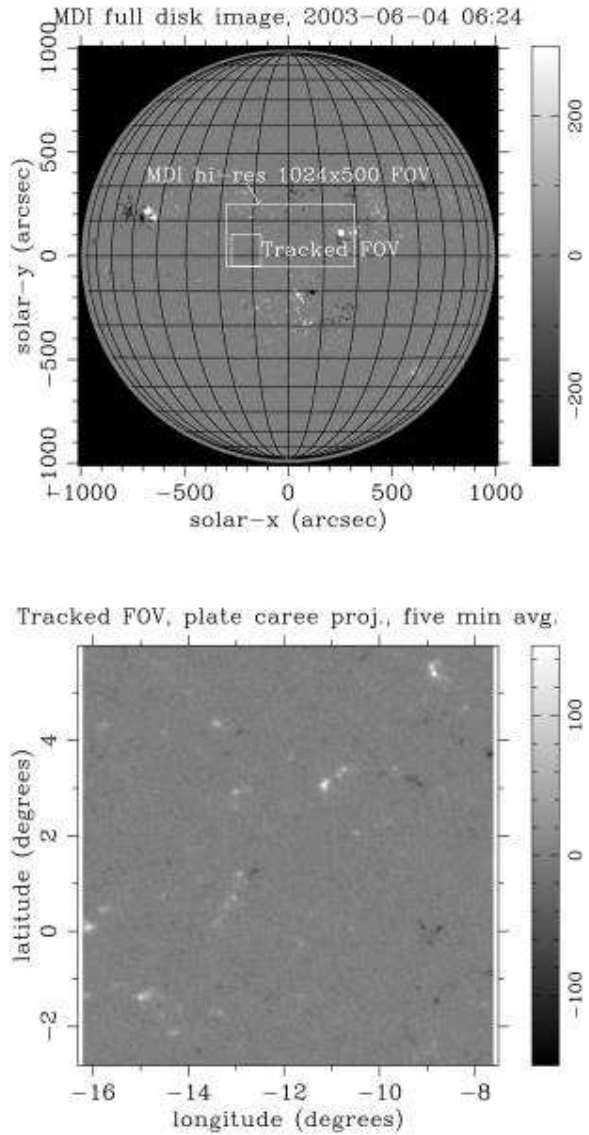}}\end{center}

\caption{\label{fig:context}The tracked field of view, in context. Note that
the edge of the grey circle in the MDI full-disk image is not the
limb of the Sun, it is a crop radius for the instrument, just outside
the limb.}
\end{figure}

\begin{figure}[!t]
\begin{center}\center{\includegraphics[%
  width=3in,
  height=2.285in]{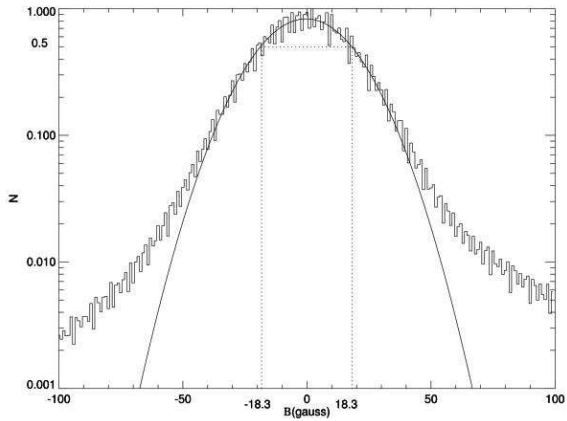}}\end{center}

\caption{\label{fig:dist-func}Distribution function of weak-field pixels
in the test dataset, a sequence of five-minute average high resolution
magnetograms from SOHO/MDI. A Gaussian fit to the low-valued pixels
(presumed to be noise) is shown. The measured standard deviation ($\sigma$)
of the images is 18.3 Gauss.}
\end{figure}

\subsection{Description of the Dataset}

We analyzed a sequence of 600 one-minute-cadence MDI high resolution
quiet-sun images from 2003 June 04, beginning at 05:43 UT. The images
were resampled into heliographic longitude/latitude coordinates (\emph{plate
caree} projection) using an orthographic model of the solar image,
as shown in Figure \ref{fig:context}. The reprojection used the ANA
language resampling tools by R. Shine (1999, personal communication). The tracked
images were 300x300 pixels and ran over the range $-16.3^{\circ}$
--- $-7.3^{\circ}$ in longitude and $-2.8^{\circ}$---$6.2^{\circ}$
in latitude. This scale slightly enlarged the images, to a pixel size
of 0.03 heliocentric degrees (about 0.48 observer arc seconds at disk
center). Images were derotated to the central time in the data sequence,
using the \citet{Snodgrass1983} synodic differential rotation curve
and rigid-body rotation at a latitude of $14^{\circ}$. These derotated,
plate caree images were averaged together in blocks of five minutes
each to reduce shot noise and P-mode interference. The noise level
was determined by fitting a Gaussian profile to the weak portion of
the pixel strength distribution curve (Figure \ref{fig:dist-func}).
The width ($\sigma$) of the best fit Gaussian profile was 18.3 Gauss,
which should be taken as the sum of all incoherent noise components
(principally shot noise, granulation, and P-mode leakage). Small frame-to-frame
offsets of the zero point (presumably due to variations in the instrument's
exposure time) were found by measuring the offset from zero for the
best-fit Gaussian, and removed by subtraction from each frame.

\subsection{Feature size distribution}

\begin{figure*}[tb]
\center{\includegraphics[%
  width=6in,
  height=2.441in]{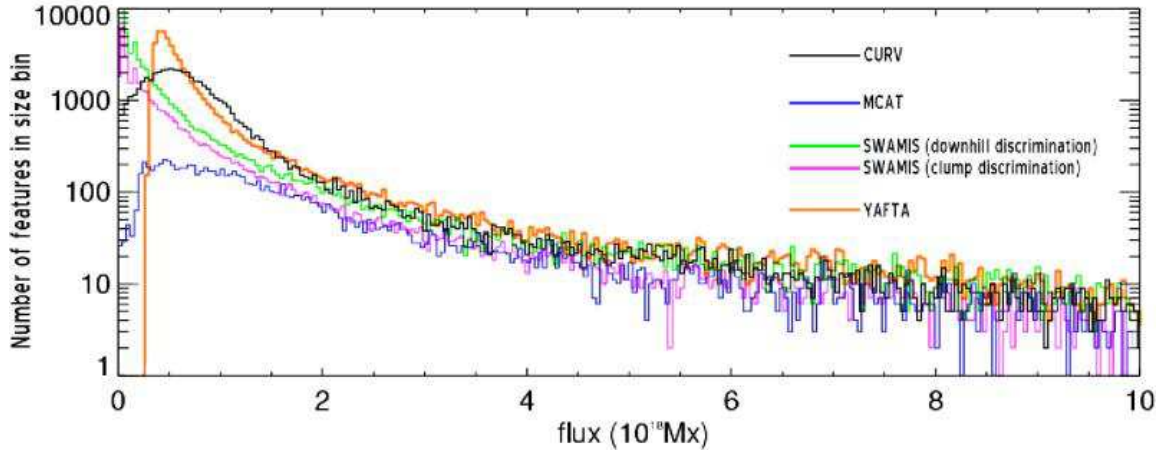}}

\caption{\label{fig:sizes}Network feature flux distributions as derived by
the five algorithms we compared in \S3. See text for full discussion. }
\end{figure*}

The simplest comparison to make across codes is distribution of fluxes
of detected magnetic features. Figure \ref{fig:sizes} shows the results
of applying all four of our codes (with two different identification
techniques for SWAMIS) to the same data. The five different techniques
yield obviously different flux distributions for the network; here
we discuss the features in the plots and the differences between them.
The plots all have the same height scale and the same bin size, so
the histograms are directly comparable. All the codes exhibit high
and low threshold behaviors that are discussed below; but it should
be immediately apparent by inspection of Figure \ref{fig:sizes} that
the codes diverge at the small end of the flux spectrum, achieving
a moderately good agreement in slope only for flux concentrations
larger than about $2\times10^{18}$~Mx. All four methods produce a
slope of about $-0.35\pm0.05$ decades per $10^{18}$~Mx, corresponding
to an e-folding width of about $1.2\times10^{18}$~Mx in the distribution.

All of the codes display a weak-feature threshold effect (false
negatives) as small features that are close to the noise floor are
eliminated by the discriminators. YAFTA and MCAT show the strongest
threshold effects, because they rely on a combination of minimum
strength and minimum feature size in each frame to eliminate
false-positives from the discrimination step. The YAFTA threshold is
particularly abrupt because the initial detection discriminator uses
no hysteresis, so that all detected features must have a minimum
number of pixels with a minimum amount of flux per pixel. MCAT's
threshold is softer because the hysteresis feature of the
discriminator allows weaker pixels to be detected around a strong
core; the dearth of very weak features is due to the combination
lifetime-and-strength requirement, which removes many weak features
that are detected by the other codes. CURV shows a still softer turnover
and threshold because the CURV discriminator does not rely on a high
threshold value in any one pixel to trigger detection.  The turnover
at about $1\times10^{18}$~Mx reflects the geometrical factor of 3 that
is applied to CURV measurements, together with the requirement for a
9-pixel concave-down region.  SWAMIS shows no obvious threshold at all
because its recursive temporal hysteresis admits many features that
have no strong pixels in a particular frame: provided that a feature
has a single strong pixel at any point in its lifetime, all of its
pixels are subjected to the weaker threshold.

The disagreement between the different codes on the weak feature distribution
is telling: it is difficult to distinguish reliably the flux distribution
of magnetic features that are smaller than about $10^{18}$ Mx in
strength even with time-averaged and conditioned MDI data. The MDI
hi-res $1\ \sigma$ detection threshold in our time-averaged data
is about $2.2\times10^{16}$ Mx, corresponding to a single pixel with
an 18 Gauss signal (Figure \ref{fig:dist-func}); features with less
than $50\times$ this much flux are not reliably detected across methods.

The greater weak-feature counts of CURV and YAFTA compared with SWAMIS
and MCAT do not necessarily correspond to greater sensitivity: our
data set was not controlled for false positives. When characterizing
a code for weak feature sensitivity, one should use noise injection
null techniques to identify false positive rates. Likewise, despite
the lack of obvious threshold the SWAMIS weak-feature distribution
curve should not be trusted below about $1\times10^{18}$ Mx because the
hysteresis requirement may reject many transient weak features that
never happen to achieve the flux density required to trip the high
threshold.

In the moderate-strength feature range of $2-5\times10^{18}$ Mx,
all four of YAFTA, MCAT, SWAMIS/downhill, and SWAMIS/clump are in reasonably
good agreement, with the main difference being between the downhill-like
codes (CURV, YAFTA, and SWAMIS/downhill) and the clumping codes (MCAT,
SWAMIS/clump). The difference is due to the segmentation of large
features into several smaller ones, giving the downhill-like codes
slightly more small features and slightly fewer large ones.

The different codes disagree substantially on slope of the flux
distribution curve in two different regions.  Below about
$1\times10^{18}$~Mx , the codes diverge strongly  in feature counts and all have 
distribution features that might serve to indicate a transition to noise-dominated
numbers: the slopes change in all the codes, and codes with thresholds of various
sorts exhibit turnover behaviors due to those thresholds.

In the small-feature range $1-1.5 \times 10^{18}$~Mx, each individual curve
has no clear indication that the data are becoming unreliable, but the different
detection schemes give divergent results.  CURV and YAFTA find more small features than
MCAT or SWAMIS in this range, due to a combination of noise and
higher sensitivity.  MCAT, which has the most stringent noise-elimination steps in 
the detection code, detects signifficantly fewer features in this size range, 
yielding a lower slope.

In the window of $~1.5-7\times10^{18}$~Mx, all three codes agree on
the slope of $-0.28\pm0.03\ \textrm{decade}^{-1}$, or an e-folding width of
$1.55\pm0.15\ \times\ 10^{18}$~Mx. The downhill-like methods find the
steeper limit and the clumping methods find the shallower
limit. Features in this size range are strong enough to be detected by
all three discriminators but not so large that the differences between
the large-scale behavior of the three codes is important.  We conclude
that features in this size range are easily detectible with MDI and
results that use this feature size range are robust against small
changes in detection technique.

The large-feature performance of the codes varies slightly across
algorithm, though all four algorithms are in rough agreement below
$10^{19}$~Mx.  At higher values the feature counts are too low to
provide good statistics, but general comments are possible.  The CURV
discriminator tends to break up large features into multiple small
features, and very large features tend to have wider wings than the
Gaussian profile that is assumed by CURV, slightly lowering the number
of detections well above $10^{19}$~Mx. YAFTA and SWAMIS/downhill also
tend to break up very large concentrations of flux into multiple
features, but that effect is not as strongly apparent. SWAMIS/clump 
and YAFTA agree quite well on the flux distribution from the YAFTA
threshold to several $\times10^{19}$Mx.  In this size range, results appear
reproducible but care is needed when inferring physical values from 
tracking results as the results appear dependent on the method used to 
identify individual features.

\begin{figure*}
\center{\includegraphics[%
  width=6in,
  height=2.441in]{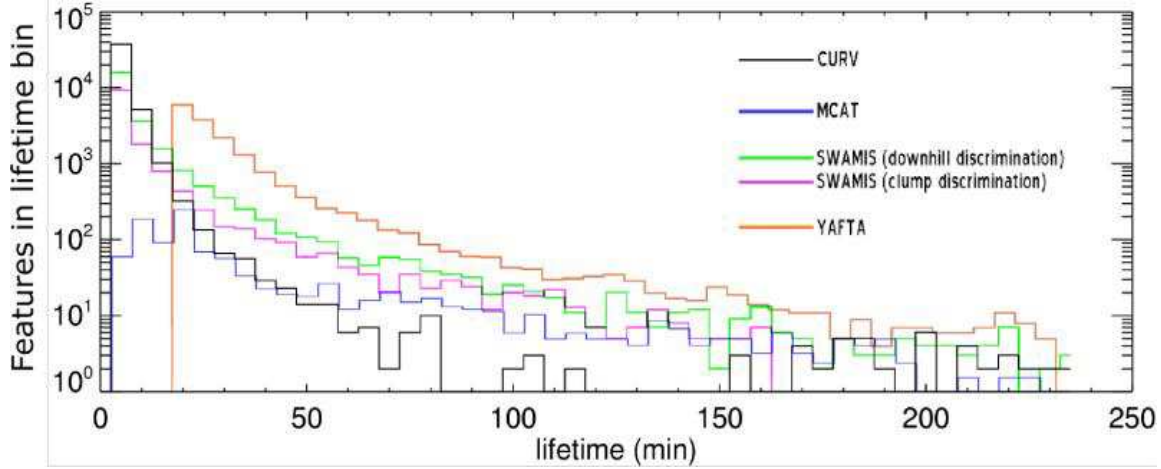}}

\caption{\label{fig:lifetime} Measured feature lifetime is strongly
dependent on tracking technique, as seen by comparing lifetime histograms
derived from each of the authors' separate tracking codes.  See text
for full discussion.}
\end{figure*}

\subsection{Feature Lifetimes}

Feature lifetime is strongly affected by the feature-association step
of the codes and hence cross-code comparison is important to identify
how reproducible that step is.  Feature lifetime is also important to
the solar physics: it is used as important measure of flux turnover
rate (e.g. \citet{Hagenaar2003}), although some physical effects other than 
flux turnover can affect it.  By comparing our codes we obtain a measure
of the reliability of feature lifetime measurements in the literature.

Several potential effects can introduce errors into flux turnover
rates measured with feature tracking codes.  In particular,
fragmentations and mergers of like-signed features cause end-of-life
events, while the associated magnetic flux survives.  Similarly,
fluctuations in the total flux or the area of a small feature cause
many birth and death events.  These effects tend to shorten the
measured lifespan of features, causing an apparent (but not real)
increase in the turnover rate of magnetic flux.

Similarly, all of our codes observe many features that are born and/or
die in a way that does not apparently conserve magnetic flux; these
events may be due to asymmetries in the field strength of small
bipoles, or due to statistical fluctuation in a collection of very
small, unresolved concentrations of magnetic flux.  If the latter is
true, then the individual unresolved concentrations that make up a
feature must have much shorter lifespans than the resolvable feature,
causing an apparent (but not real) decrease in the turnover rate of
magnetic flux.

Figure \ref{fig:lifetime} shows a histogram plot of feature lifetime
from each of our codes. The codes agree on the slope (but not the
value) of the lifetime histogram for a narrow range of lifetimes
between 20-50 minutes.  Roughly 75\% of features found by SWAMIS in
this range of lifetimes are in the $2-6\times 10^{18}$~Mx size range
in which the codes agree on feature counts, suggesting that this
region of slope agreement is similar to the region along the size
axis: features in this population are high enough above the noise
floor to be readily detectible but not so large nor long-lived that
geometrical effects fool the different tracking algorithms.

The differences between the curves are entirely due to differences in
the algorithms of the codes, as we examined identical data.  MCAT
requires longevity of more than two frames (10 minutes) in the
identification step; the small number of one-frame features are due to
fragmentations (features that fragment from an existing feature, then
disappear one frame later). SWAMIS and CURV use much weaker longevity
requirements, and therefore detect similar numbers of short-lived
features.  YAFTA detects many more short-lived features than the other
codes, in part because of a lower detection threshold (no detection
hysteresis was used for this data set) but does not include any
features with less than a 4-frame (20 minute) lifetime.

CURV, alone of all the codes, shows a minimum in the lifetime
histogram followed by a slight rise in the 150-200 minute range.  The
population in the rise consists of nearly 100 structures, enough to be
statistically significant compared to just 9 features features found
by CURV in this data set with lifetimes between 100 and 150 minutes.
It is not clear whether this is an unusual statistical event or a 
quirk of the CURV association scheme.

The main conclusion to draw from this comparison is that feature
lifetimes are extremely difficult to measure with tracking codes; in
consequence, average magnetic lifetime results from magnetic tracking
of arcsecond-scale data should be considered weak.  We will address
the nuances of lifetime measurement, and its relevance to physical
parameters such as magnetic turnover time and heating rate, in a later
article in this series.

\section{\label{sec:Discussion}Discussion}

Each of the techniques we considered has advantages for a particular
regime of fragment size and strength relative to the noise floor of
the instrument. Here, we discuss the tradeoffs of the different detection
schemes. The main differences between our codes lay in the discrimination
and feature-identification steps, which are discussed separately.

\subsection{Discrimination}

The main problem faced by tracking discriminators is the huge number
of statistically independent samples across an image sequence dataset.
The simplest discriminator is a threshold trigger; while threshold
triggers are inadequate for many tasks when used alone, they form the
basis of every discrimination algorithm. The three main ways we
improved upon simple threshold triggering were curvature sensing
(CURV), hysteresis (MCAT, SWAMIS), and post-discrimination filtering
for feature size and longevity (all codes). Simple trigger
discrimination is useful mainly where the signal-to-noise ratio is
overwhelmingly large.  One code in our study (YAFTA) was optimized for
strong field detections and used simple trigger discrimination, although
subsequent versions of YAFTA include the ability to use hysteresis.

Curvature sensing as implemented in CURV has the advantage that, when
combined with a threshold trigger, it applies five statistically
independent threshold tests to each pixel, significantly reducing the
false-positive rate. CURV rejects features whose convex cores are
smaller than 9 pixels. Including the effects of smoothing in the
preprocessing steps, which leave granulation as the dominant source of
noise, there are about 12 statistically independent tests (of 45 total
conditions) required to detect a particular feature. By contrast, a
direct trigger yields only about three statistically independent tests
with the same size threshold. Curvature discrimination permits a
detection threshold much closer to the noise floor than would
otherwise be possible, which in turn should make curvature
discrimination rather sensitive to weak concentrations of magnetic
flux.

The disadvantage of curvature discrimination is that it only finds the
convex core of a magnetic feature. This is addressed by
\citet{Hagenaar1999} via a simple scaling: they find that for a large
variety of near-Gaussian distributions the convex core is about 1/3 of
the total flux in the feature, and scale accordingly. This works well
for small features near the resolution limit of the observations, but
not as well for larger features, which are observed to have flatter
profiles than a Gaussian. Large concentrations of flux typically have
several local maxima, and the total flux may be over- or
underestimated by the assumption of a simple Gaussian shape, depending
on the actual morphology of the feature.

Hysteresis is a simple way of reducing the false positive rate of
threshold-trigger discrimination. Pixels are compared against
different trigger thresholds depending on whether they are isolated or
adjacent to other detected pixels. Both MCAT and SWAMIS use a
recursive-hysteresis scheme that starts with a simple threshold
scheme, and then dilates detected pixels using a lower threshold. Such
schemes eliminate many false positives due to the background noise
floor, and detect the full extent and shape of large features. The
drawback is that weak features are only detected if they have at least
one 'seed' pixel that is stronger than the high threshold. SWAMIS
further allows dilation along the time axis, so that weak features are
detected if at some point in their lifetime they have a single strong
pixel; but even so, many transient weak features that are visible to
the eye go undetected for lack of a single strong pixel.

\subsection{Feature Identification}

Here, we contrast the two principal dilation strategies of the codes:
downhill and clumping dilation from local maxima. The distinction
between these strategies is academic within CURV, as the curvature-based
discriminator provides well-separated loci around each local maximum:
the final detected loci are the same regardless of dilation method.
MCAT, SWAMIS, and YAFTA can dilate using clumping, and SWAMIS and
YAFTA can dilate using the downhill technique.

Both the downhill and clumping techniques, together with hysteretic
thresholding, do better than curvature at identifying the size and
shape of mid-sized magnetic features in the several arc second size
range. In this size range the shape of individual features varies
considerably, though most features still have but one local maximum
in the MDI data that we tracked; under these conditions, both dilation
techniques do about as well as one another and both measure the flux
of individual features with more precision than CURV (which uses a
simple geometric factor to estimate the flux in the wings of the structure).

\begin{figure}
\center{\includegraphics[%
  width=3in,
  height=1.14in]{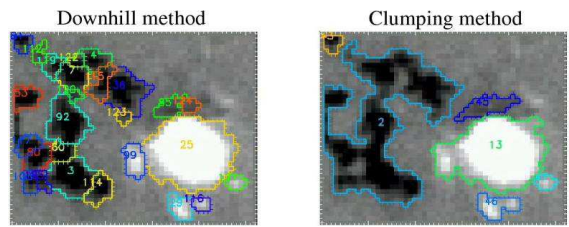}}

\caption{\label{fig:methods}Effect of feature identification technique: a
pathological case. Clumping (right) is less sensitive to noise in
weak features than is downhill dilation (left), but can lead to counterintuitive
results in active regions as large, irregular patches of flux are
identified as a single object.}
\end{figure}

Large scale structures that are more than about 15 arc seconds across
yield stronger differences between the downhill and clumping techniques,
as illustrated in Figure \ref{fig:methods}. The downhill technique
does a better job at tracking substructure of large, extended objects
such as plage and active region fields, but at the expense of more
noise susceptibility. Because small amounts of noise can produce transient
local maxima in a large extended feature, the downhill technique is
susceptible to the \emph{swiss cheese problem} in which a single large
clump of flux with no strong local maximum can be oscillate between
being detected as one or several separate features. If lifetime filtering
is being applied to the detected-feature list, then large holes may
appear in the detected feature, giving it the appearance of an irregular
block of Ementhaler cheese. Furthermore, downhill detection alone
tends to miss very large concentrations of flux, treating them as
a collection of smaller features: while this is desirable for tracking
the motion of the solar surface, it is not desirable when measuring
the statistics of strength or size of magnetic features.

We discuss the tradeoff between these detection techniques, neither of
which is perfect, in \S\ref{sec:recommendations}, below.

\subsection{\label{sub:Cross-frame-feature-association}Cross-frame feature association}

All of the codes we compared use essentially the same cross-frame
association strategy of finding the association map that maximizes
overlap between features in adjacent frames, as described in
\S\ref{sub:Feature-association}.  In practice, most features in most
frames overlap with exactly one feature in the following and adjacent
frames, so variations in the type of overlap (e.g. number of pixels
vs. amount of flux) or of permissiveness of overlap (e.g. including
pixels nearby each feature as part of the feature itself, for purposes
of finding overlap) only affect the {}``edge cases'' in which multiple
magnetic features are interacting, or in which a single feature is
moving rapidly.

Overlap-style algorithms such as described in
\S\ref{sub:Feature-association}\ are quite robust for associating
features with the properties $\forall
i,j:\left|\overrightarrow{x_{i}}-\overrightarrow{x_{j}}\right|\gtrsim
r_{i}$ and $\forall
i:\left|\Delta\overrightarrow{x_{i}}\right|\lesssim r_{i}/2$, where
$i$ and $j$ are indices across features, $\overrightarrow{x_{i}}$ is
the centroid location of feature $i$, $r_{i}$ is the typical radius of
feature $i$, and $\overrightarrow{\Delta x_{i}}$ is the displacement
vector of feature $i$ across frames. Fast moving features with
$\left|\Delta\overrightarrow{x_{i}}\right|\gtrsim r_{i}/2$ become
subject to the \emph{mistaken identity problem}, where they are
identified as a different feature in different frames. The mistaken
identity problem affects statistical feature-lifetime and feature
history results even if only a very few features are subject to it,
because a single fast-moving feature may register as a very large
number of separate magnetic features. The only reliable way to beat
the mistaken identity problem is to use high enough time resolution in
the data. Marginal data in which the fastest moving features have
$\Delta\overrightarrow{x_{i}}\thickapprox r_{i}$ may be improved by
interpolating interstitial frames, but wider separations cannot be
helped by that method. In practice, we found by visual inspection that
12-minute final effective cadence with direct boxcar averaging was not
sufficient to avoid the mistaken identity problem for the fastest
moving features at the MDI {}``full disk'' resolution (1.4 Mm pixels
at Sun center), accounting for \textasciitilde{}0.5\% of features in a
given frame and perhaps 5\% of total features identified by SWAMIS,
but that 12 minute final effective cadence with anti-aliasing in the
time direction (time-weighted averaging of 1 minute cadence
magnetograms, with a 12 minute FWHM Gaussian weighting profile)
eliminates virtually all cases of mistaken identity.

We considered, but did not implement, various methods to reduce the
mistaken identity problem in cases where high enough cadence data are
not available. Promising directions to try include linear
extrapolation of feature location from last associated location before
the overlap calculation, with or without backtracking on the time
axis; dilation with size checking; and simulated annealing of feature
association.  We suspect that all such algorithms are likely to
include more faulty associations than does direct overlap.

\section{Recommendations on Accepted Practice\label{sec:recommendations}}

To enable meaningful comparison and reproducibility of tracking results
across research groups, we recommend the following techniques as appropriate
for most applications of magnetic feature tracking.

\subsection{Data preprocessing}

While preprocessing of data is not technically a part of feature tracking,
preprocessing can affect the statistics of image tracking and therefore
warrants mention here. We discuss despiking, time averaging, and resampling
into a desired coordinate system.

\paragraph{Despiking}

A brief note on space-based magnetograms is in order: \emph{SOHO}/MDI
is, and presumably SDO/HMI will be, susceptible to cosmic ray impacts.
A typical MDI {}``full-disk'' magnetogram has evidence of cosmic
ray impacts in $~10^{2}$ pixels, so 5-10 minute averages may have
as many as $~10^{3}$ bad pixels caused by cosmic rays. The cosmic
rays are not saturated in the images, and may have either negative-going
or positive-going direction. These cosmic rays can skew the size,
strength, and lifetime statistics of small, short-lived features if
not considered. We recommend either despiking sequences of space-based
magnetograms with a second time derivative technique such as ZSPIKE
(\citealt{DeForest2004b}), or imposing a lifetime threshold on detected
features to limit the effects of cosmic rays.

\paragraph{Time and spatial averaging}

Time averaging of images is useful as a preparatory step to reduce
noise in the magnetograms and to smooth features for better association
across frames. There are several sources of noise in currently available
magnetograms, with different statistics for each source; we discuss
them briefly here, as the noise characteristics of averaged data sets
hold a complex relationship to the noise characteristics of individual
frames.

Most magnetographs are photon-limited, so that there is an approximately
Gaussian distribution noise source of \emph{photon noise} associated
with photon counting statistics in each pixel of each magnetogram.
Each pixel contains an independent sample of this noise source. Magnetographs
such as MDI that assemble multiple exposures are subject to \emph{shutter
noise}, which results from very slightly different exposure times
across each independent exposure used to produce the magnetogram:
shutter noise is an approximately Gaussian distribution noise source
that is added to the common mode of all pixels across each image.
Finally, solar evolution (and, for ground-based telescopes, seeing
effects) across the time of assembly of the magnetogram induces an
additional noise source, \emph{evolution noise} that is dominated
by the evolution of granules. Granulation, and the associated evolution
noise, has about one independent sample every five minutes, per square
megameter of solar surface area.

Individual MDI magnetograms have about equal amounts of photon and
evolution noise. Because the photon noise is independently sampled
in each image, averages of more than about five minutes of magnetic
data tend to be dominated by evolution noise, which is attenuated
much more slowly by further averaging. 

Anti-aliased time averaging, using overlapping Gaussian or Hanning
windows in the time domain, is preferable to simple boxcar averaging,
which has frequency sidelobes that allow more noise to enter the data.

\paragraph{Image resampling}

Image sequences are typically resampled to remove the solar rotation
and perspective \emph{a priori.} For the present work, we derotated
and prepared time averages of MDI magnetograms using a simple interpolation
scheme into plate caree coordinates. This scheme follows current common
practice but is not recommended: it fails to preserve small-scale
feature statistics in two important ways.

First, the plate caree ({}``lon/lat'') map projection is \emph{non-authalic}:
a feature of unit area on the surface of the Sun may have different
areas in plate caree coordinates, depending on its latitude. To avoid
skewing the statistics of flux content, authors should use an authalic
(equal-area) projection to prepare the data before tracking. The area
of a feature in the plate caree projection is scaled by a factor of
secant(latitude). A simple way to compensate is to scale the vertical
or horizontal scale by cos(latitude) at each point. Scaling the vertical
axis by cos(latitude) yields the common \emph{sin-lat cylindrical
projection}, so named because integrating the scale factor $y'_{map}~cos(lat)$
yields $y_{map}~sin(lat)$. Scaling the horizontal axis yields the
\emph{sinusoidal projection}. Other useful authalic choices include
the Hammer/Aitoff elliptical projection used by the cosmology community
and Lambert's azimuthal equal-area projection, which minimizes linear
distortion near the origin. Many useful projections have been cataloged
by \citet{Snyder1987}.

Secondly, linear interpolation leaves much to be desired as a resampling
method, skewing (among other things) the noise profile of individual
pixels and potentially introducing large amounts of distortion into
the statistics of small features. A statistically sound, photometrically
accurate resampling method, relying on spatially variable sampling
filters, has been described by \citet{DeForest2004}; that or similar
techniques are recommended for preparing data for survey applications.

For virtually every application of tracking, it is important to compensate
by rigid rotation based on the differential rotation speed at a particular
point in the field of view, and not by differentially rotating every
pixel in the image independently. The former preserves the actual
evolving spatial structures in question; the latter only preserves
the plasma reference frame at the start of the observing run.

\subsection{Feature discrimination \& identification}

We recommend combining the three methods of feature detection. Standard
codes should use a dual-discriminator scheme for detection: an initial
convex-core discrimination as in CURV, followed by dilation
to a low noise threshold. This combination takes best advantage of
the extra discrimination afforded by the convex core technique, while
eliminating some of the difficulties of identifying oddly-shaped and
large features. 

Feature identification should use the downhill method to avoid pathologies
of the clumping technique, particularly when used for motion tracking
and to identify interacting magnetic features; but for applications
where larger clusters of flux are important, we recommend keeping
track of groups of touching or nearby features according to a clumping
algorithm. Groups of mutually touching features are the same loci
as would be identified by a direct clumping scheme, but tracking individual
peaks within the group affords better localization of the magnetic
flux that makes up the feature(s). This can be accomplished either
by maintaining a table of mutually touching features or by using a
dual-labeling scheme at the feature-identification step.

\subsection{Feature association}

For best general purpose utility, we recommend a flux-weighted maximum
overlap method of association between frames, as is currently used
by SWAMIS; for example, in cases of associative conflict such as Figure
\ref{fig:assoc}, regions B and C would be associated as identical,
region A would be classified as dying by merger into B/C, and region
D would be classified as originating by fragmentation from B/C. For
analyses that require feature identification, it is important to ensure
that the cadence is sufficient to allow associated features in adjacent
frames to overlap. While more sophisticated motion-correlation algorithms
are in principle feasible, they add complexity and fallibility that
is not necessary provided that the data have high enough cadence.

It is notable that no local overlap algorithm agrees with a human
observer in all cases, as human observers use more information than
strict overlap -- including something like a predictor/corrector position
algorithm. Maximum overlap works well in the case where the motion
of all features is small compared to their width divided by the time
step. If the time step is too long, small features can move more than
their diameter in a single frame, leading to the mistaken identity
problem, where a single visually identifiable feature frequently
changes its identity in the tracked data. In such cases, one can (i)
use faster frame rates, (ii) generate dilated feature masks for association,
(iii) use linear location extrapolation to account for the large interframe
motion, and/or (iv) use a minimum-distance criterion rather than maximum
overlap.

\subsection{Feature tabulation}

When tabulating feature histories, we recommend that the following
minimum information be kept for each feature, and for each frame for
which a particular feature exists: area (A), total flux ($\Phi$),
flux-weighted average location (\emph{x,y}), and flux-weighted quadrupole
moments ($<\Phi^{2}dx^{2}>,<\Phi^{2}dy^{2}>,<\Phi^{2}dxdy>$), for
a total of 7 numerical quantities per feature per frame. The quadrupole
moments, in particular, summarize tersely and simply the shape of
the feature, and the features detected by the downhill dilation method
tend to be simple shapes that are readily described with the quadrupole
moment set. Quantities may be kept in physical or image units (e.g.
km or pixels). Quantities which we recommend avoiding are: pixel value
maximum and variance, which depend on resolution and phase of the
underlying feature relative to the pixel grid; and non-weighted average
location, because it is more dependent on noise-dominated pixels at
the feature's edge than is the flux-weighted average location.

\subsection{Event identification}

Several of the scientific applications of tracking require classifying
the origin and demise of each feature based on visual heuristics for
the underlying physics. Useful event classification requires characterizing
the geometry and manner of change of nearby features. Event classification
is a rich topic that is not fully discussed in this paper; however,
we make some brief recommendations.

We recommend classifying origin events into four categories: (i) \emph{isolated
appearance}, in which a particular feature appears in the absence
of interaction with surrounding detected features; (ii) \emph{balanced
emergence}, in which a bipolar, approximately balanced pair of features
appear together in nearly the same location at nearly the same time;
(iii) \emph{unbalanced emergence}, in which a new feature appears
next to a pre-existing, opposite sign feature in a nearly flux-conserving
manner; and (iv) \emph{fragmentation} (or splitting), in which a single
pre-existing feature breaks up into multiple smaller features in a
nearly flux-conserving manner. Demise events should be classified
in the exact same way as origin events, in a time reversed sense:
(i) \emph{isolated disappearance}; (ii) \emph{balanced cancellation};
(iii) \emph{unbalanced cancellation}; and (iv) \emph{merging}. For
both origin and demise events, (i) is the only recognized case that
apparently violates conservation of flux; (ii) corresponds to isolated
passage through the photosphere of a magnetic loop; and (iv) represents
reshuffling of existing flux. For completeness, event identification
software should also maintain a \emph{complex} class for events which
cannot be classified easily into the above four groups, including
such events as isolated asymmetric emergence that violate conservation
of magnetic flux.

It is important to understand that this is a \emph{visual} classification
scheme, to be more fully developed in future work. Interpretation
of these visual events in terms of physical mechanisms is neither
straightforward nor obvious. For example, appearance events may or
may not correspond to new flux on the solar surface.

Physical modeling of feature behavior requires some care.  In particular, 
only some emergence events (bipolar emergence) appear to be due to flux
tubes that emerge from below the surface of the Sun
(\citealt{HarveyMartin1973,HarveyKL1993,Chae2001}).  Such events should give
rise to two oppositely signed magnetic features that grow together and
separate in a divergent surface flow
(\citealt{Hagenaar2001,Hagenaar2003,SimonTitleWeiss2001}), and the
origin detection code in SWAMIS and in CURV was originally intended to
identify such events. However, proportionally few magnetic
features are observed to originate with this \emph{balanced emergence}
mechanism.  New small features can also form by \emph{fragmentation}
of pre-existing large features into like-signed fragments; this
process is also called \emph{calving} if the new feature is small
compared to the surviving feature. Furthermore, many features simply
\emph{appear}, without any surrounding flux at all or in ways that
appear to violate flux conservation. The nature of these appearances
-- whether \emph{coalescence} of existing weak flux or
\emph{unbalanced emergence} with one large, weak-field pole and one
small, strong-field pole, will be considered in detail in Paper II of
this series.

Results using our recommended classification scheme should be presented
together with a notation describing what criteria are used to detect
balanced changes in the flux of interacting features. Event classification
results can be quite different, for example, if the changes in the flux
of two interacting features are considered {}``approximately balanced''
if they merely have opposite sign, or if they must agree within,
say, 10\%.

\section{\label{sec:Conclusions}Conclusions}

We have compared four magnetic feature tracking codes by applying
them to the same preprocessed set of magnetic data. Feature tracking
code output is sensitive to a variety of decisions that are made during
development, and this sensitivity is a reason why it has historically
been difficult to reproduce results obtained by feature tracking:
it is crucial to explain exactly what algorithm is being used. In
particular, codes that were designed for one regime of study (e.g.
very small intranetwork flux concentrations or very large, strong
features) should not be applied to different regimes of detection
without careful study, and all discrimination and association techniques
need to be lain out exactly as performed.

The difficulty of reproducing apparently simple results in feature
tracking appears to stem both from the complicated, noisy nature of
the magnetograph data and from the complexity of the underlying
structures.  The solar magnetic field is not divided into well
separated, strongly magnetized features; rather, there is a continuum
of feature sizes due to the clustering behavior of the field across
scales, in keeping with the concept of \emph{magnetochemistry}
outlined by \citet{Schrijver1997}.  Bulk summary characteristics such
as the lifetime of individual features or the size distribution of the
features depend strongly both on the instrument being used to image
the magnetic field and on threshold and related decisions made during
code development.

All of our codes agree reasonably well on important summary
characteristics in a particular circumscribed range of scales and
lifetimes, indicating that there is an underlying pattern to be
measured; but the region of agreement (which we take to be the range
of valid measurement using the tracking codes) is much smaller than
might be surmised from cursory analysis of the output of any one
algorithm.  We conclude that particular care must be used when
interpreting magnetic tracking results, which are often much weaker
than might be surmised given the apparent clarity of solar magnetic
features in magnetogram sequences.

In particular, we find that the magnetic turnover time, perhaps the most 
accessible summary result to come out of magnetic tracking studies, is
also perhaps the weakest result to come out of magnetic tracking studies.
Average feature lifetimes are only weakly related to magnetic turnover time
in the best of circumstances, and we have found that average lifetime 
measurements are strongly dependent on the code being used to perform
the measurement.  

By comparing and contrasting the algorithms of our four separate
codes, we have determined why the they produce different results for
the flux distribution in quiet sun, and evaluated under what
circumstances each technique performs best. Further, we have made
recommendations about how to improve feature detection and
reproducibility in feature tracking for future work. To aid that work,
all four of our codes are being made available to the scientific
community in source-code form via \emph{solarsoft. }

Specific physical problems such as flux emergence and cancellation,
diffusion of active region flux and plage formation, and feature
lifetime, will be covered in more detail in future papers in this
series.

\bibliographystyle{plainnat}

\clearpage\addcontentsline{toc}{chapter}{\bibname}

\begin{thebibliography}{27}
\providecommand{\natexlab}[1]{#1}
\providecommand{\url}[1]{\texttt{#1}}
\expandafter\ifx\csname urlstyle\endcsname\relax
  \providecommand{\doi}[1]{doi: #1}\else
  \providecommand{\doi}{doi: \begingroup \urlstyle{rm}\Url}\fi

\bibitem[{Chae} et~al.(2001)]{Chae2001}
{Chae}, J. et~al. 2001,
\newblock \apj, 548, 497

\bibitem[{DeForest}(2004)]{DeForest2004}
{DeForest}, C.~E. 2004,
\newblock \solphys, 219, 3

\bibitem[{DeForest} and {Lamb}(2004)]{DeForestLamb2004}
{DeForest}, C.~E. and {Lamb}, D.~A. 2004,
\newblock BAAS, 204

\bibitem[{DeForest} (2004)]{DeForest2004b}
{DeForest}, C.~E. 2004,
\newblock \apj, 617, L89 

\bibitem[{Fossum} and {Carlsson}(2004)]{Fossum2004}
{Fossum}, A. and {Carlsson}, M. 2004,
\newblock ESA SP-547, 125

\bibitem[{Hagenaar}(2001)]{Hagenaar2001}
{Hagenaar}, H.~J. 2001,
\newblock \apj, 555, 448

\bibitem[{Hagenaar} et~al.(1999)]{Hagenaar1999}
{Hagenaar} H.~J. et~al. 1999,
\newblock \apj, 511, 932

\bibitem[{Hagenaar} et~al.(2003)]{Hagenaar2003}
{Hagenaar}, H.~J. , {Schrijver}, C.~J., and {Title}, A.~M. 
\newblock \apj, 584, 1107

\bibitem[{Handy} et~al. (1999)]{Handy1999}
{Handy}, B.~N. et~al. 1999
\newblock \solphys, 187, 229

\bibitem[{Harvey}(1993)]{HarveyKL1993}
{Harvey}, K.~L. 1993
\newblock Ph.D.~Thesis, Univ. of Utrecht

\bibitem[{Harvey} and {Martin}(1973)]{HarveyMartin1973}
{Harvey}, K.~L., and {Martin}, S.~F. 1973,
\newblock \solphys, 32, 389

\bibitem[{Lamb} and {Deforest}(2003)]{LambDeForest2003}
{Lamb}, D.~A. and {Deforest}, C.~E. 2003,
\newblock AGU FMA, B530

\bibitem[{Leibacher}(1995)]{Leibacher1995}
{Leibacher}J.~W.~et~al.  1995,
\newblock ASP Conf. Ser., 76, 381

\bibitem[{Leibacher}(1999)]{Leibacher1999}
{Leibacher}, J.~W. 1999,
\newblock Adv. Sp. Res., 24, 173

\bibitem[{Parker}(1988)]{Parker1988}
{Parker}, E.~N.  1988,
\newblock \apj, 330, 474

\bibitem[{Parnell}(2002)]{Parnell2002}
{Parnell}, C.~E. 2002,
\newblock \mnras, 335, 389

\bibitem[{Parnell} and {Jupp}(2000)]{ParnellJupp2000}
{Parnell}, C.~E. and {Jupp}, P.~E 2000,
\newblock \apj, 529, 554

\bibitem[{Scherrer} et~al.(1995)]{Scherrer1995}
{Scherrer}, P.~H. et~al. 1995,
\newblock \solphys, 162, 129

\bibitem[{Schrijver} et~al.(1997)]{Schrijver1997}
{Schrijver} C.~J. et~al. 1997,
\newblock \apj, 487, 424

\bibitem[{Schwer} et~al.(2002)]{Schwer2002}
{Schwer}, K.~ et~al. 2002,
\newblock AGU FMA, C1

\bibitem[{Simmons}(1972)]{Simmons}
{Simmons}, G.~F. 1972,
\newblock McGraw-Hill

\bibitem[{Simon} et~al.(2001){Simon}, {Title}, and {Weiss}]{SimonTitleWeiss2001}
{Simon}, G.~W.,  {Title}, A.~M., and {Weiss}, N.~O. 2001,
\newblock \apj, 561, 427

\bibitem[{Snodgrass}(1983)]{Snodgrass1983}
{Snodgrass}H.~B. 1983,
\newblock \apj, 270, 288

\bibitem[Snyder(1987)]{Snyder1987}
{Snyder}, J.~P. 1987,
\newblock USGS Prof. Papers, 1395

\bibitem[{Strous} et~al.(1996)]{Strous1996}
{Strous}, L.~H. et~al. 1996,
\newblock \aap, 306, 947

\bibitem[{Welsch} and {Longcope}(2002)]{Welsch2002}
{Welsch}, B.~T. and {Longcope},D.~W. 2002,
\newblock ESA SP-505, 611

\bibitem[{Welsch} and {Longcope}(2003)]{Welsch2003}
-----------. 2003,
\newblock \apj, 588, 620

\bibitem[{Welsch} et~al.(2004){Welsch}, {Fisher}, {Abbett}, and
  {Regnier}]{Welsch2004}
{Welsch}, B.~T. et~al. 2004,
\newblock \apj, 610, 1148

\end{thebibliography}

\appendix{Glossary}

Feature tracking and magnetic observations are mature enough to have
developed a collection of commonly used terms, which unfortunately
have drifted into slightly different usage in different locations.
In an attempt to regularize terminology, we present a glossary of
commonly used terms, with their recommended definitions. Also, because
some terms are strictly observational and others imply a physical
model, we have noted which are which.

\paragraph{\emph{Object descriptions}}

\begin{description}
\item [bipole]- a pair of magnetic features of opposite sign and approximately
equal flux content, that appear to be associated (as in bipolar \textbf{emergence}).
When seen to emerge together, the poles of a bipole may be associated
observationally.
\item [ephemeral~region]- a resolved small bipole with particular properties
as described by \citet{Hagenaar2001}.
\item [feature]- a visually identifiable part of an image, such as a clump
of magnetic flux or a blob in a magnetogram. The term {}``feature''
is purely observational and is preferable to {}``flux concentration''
or {}``ephemeral region'' when describing individual visual objects
in an image. The specific definition of a feature is dependent on
both the Sun itself and the characteristics of the observing telescope.
\item [flux~concentration]- a localized cluster of magnetic flux, with
or without resolved substructure. A flux concentration may consist
of one or more magnetic features. While somewhat vague, the definition
of a flux concentration is approximately independent of observing
telescope: a flux concentration may appear as a single feature when
seen with one instrument but as several features with another.
\item [fragment]- a small piece of a larger magnetic structure, \emph{not}
a generic small bit of magnetic flux. Usage: {}``this magnetic flux
concentration is composed of many fragments'', or {}``unresolved
fragments make up this magnetic feature''. {}``Fragment'' should
not be used interchangeably with {}``feature'', as it implies that
the subject is part of a larger whole, while {}``feature'' does
not.
\item [monopole]- a lone magnetic pole (thought to be physically impossible).
\item [stenflo]- (after J. Stenflo) a tiny, strong concentration of order
$10^{17}$ Mx of flux. Usage: {}``The asymmetric formation of flux
concentrations in the network may be due to convergence of stenflos,
though Stenflo himself may object to this terminology.''
\item [unipole]- a single magnetic feature with no obvious associated feature
of the opposite sign. The photospheric boundary provides a {}``hiding
place'' for the opposing pole, so that unipoles are thought not to
be monopoles. Oppose {}``bipole'', {}``monopole''. 
\end{description}

\paragraph{\emph{Event} \emph{descriptions}}

\begin{description}
\item [appearance]- used specifically to describe the origin of a single
unipolar feature where there were none before. Appearances appear
to violate conservation of magnetic flux, but probably result from
flux hiding under the noise floor of an instrument -- so the definition
of {}``appearance'' depends on the instrument being used.
\item [asymmetric~emergence]- emergence in which the two sides of the
emerging magnetic loop of flux have quite different cross sections,
perhaps reducing the field strength of the larger leg of the loop
below the detection threshold of an instrument. This can be a physical
description of one type of feature appearance; coalescence is another
type. Note that {}``asymmetric emergence'' and {}``unbalanced emergence''
are not synonyms.
\item [balanced~emergence]- emergence in which the two final opposing-sign
features have approximately the same magnitude; this is the type of
emergence predicted by a simple model of magnetic flux tubes rising
through the photosphere. Compare {}``emergence''; contrast {}``unbalanced
emergence''.
\item [balanced~cancellation]- cancellation in which the two initial opposing-sign
features have approximately the same magnitude. Compare {}``cancellation'';
contrast {}``unbalanced cancellation''. Balanced cancellation is
the time reversal of balanced emergence.
\item [calving]- a form of fragmentation in which one of the daughter features
contains much more flux than the other, by analogy to the behavior
of icebergs. Usage: {}``This movie shows small features calving off
of the main flux concentration''. Oppose {}``splitting''; compare
{}``fragmentation''.
\item [cancellation]- the demise of a magnetic feature that collides (and
cancels) with an opposing sign feature, in such a way that flux is
approximately conserved. Compare {}``balanced cancellation'', {}``unbalanced
cancellation''; contrast {}``disappearance''.
\item [coalescence]- the collection of diffuse flux from below detection
threshold to a small, denser feature that can be detected. This may
be an example of unresolved merging. This is a physical description
of one type of feature appearance; asymmetric emergence is another
type. To avoid confusion, eschew {}``coalescence'' when describing
observational results; use {}``merging'' or {}``appearance'' instead.
\item [demise]- the end of a magnetic feature's existence.
\item [disappearance]- the end of a single, unipolar magnetic feature that
{}``fades away'' to nothing in the absence of nearby features (the
time reversal of an {}``appearance'').
\item [dispersal]- deprecated. This has been used to describe the opposite
of coalescence, the breakup of strong flux concentrations into many
fragments, and the diffusion of flux across the surface of the Sun.
It is now too ambiguous to be used clearly in most cases.
\item [emergence]- the origination of two balanced, opposing magnetic features
nearby one another in such a way that flux is approximately conserved.
This observational definition follows the common physical definition
of a loop of flux emerging from below the surface. Compare {}``balanced
emergence'', {}``unbalanced emergence''; contrast {}``appearance''.
Emergence is the time reversal of {}``cancellation''. 
\item [fragmentation]- the breakup of a single magnetic feature into at
least two like-sign features. (compare {}``splitting'', {}``calving'')
\item [merging]- the joining of two magnetic features of similar sign into
a single larger feature.
\item [splitting]- the breakup of a single magnetic feature into at least
two like-sign features, with the implication of rough flux balance
between the two daughter features. (oppose {}``calving''; compare
{}``fragmentation'').
\item [unbalanced~emergence]- emergence in which the two final opposing-sign
features have different magnitudes due to interaction with a nearby
unipolar feature. Compare {}``emergence''; contrast {}``fragmentation'',
{}``balanced emergence''. Unbalanced emergence is the time-reversal
of unbalanced cancellation.
\item [unbalanced~cancellation]- cancellation that is not complete because
one of the canceling features contains more flux than the other. Compare
{}``cancellation''; contrast {}``merging'', {}``balanced cancellation''.
\end{description}

\acknowledgements{Thanks to the SOHO/MDI team for kind use of their data, and to the
University of St. Andrews for hosting the workshop which made this
comparison possible. This work was funded by NASA's SOHO project,
the SOHO/MDI effort, NASA's SEC-GI program, the Air Force Office of
Scientific Research MURI program, and the PPARC Advanced Fellowship
program. SOHO is a project of international collaboration between
NASA and ESA.}
\end{document}